\documentclass[aps,pra,twocolumn,groupedaddress,showpacs]{revtex4-1}
\usepackage{amsmath}
\usepackage{graphicx}
\usepackage{amsfonts}
\usepackage{color}
\usepackage{upgreek}
\makeatletter
\def\captionof#1#2{{\def\@captype{#1}#2}}
\makeatother
\begin{document}

\title{Quantum transport under AC drive from the leads : A Redfield Quantum Master Equation approach} %

\author{Archak Purkayastha}
\affiliation{International centre for theoretical sciences, Tata Institute of Fundamental Research, Bangalore - 560012, India}
\author{Yonatan Dubi}
\affiliation{Department of Chemistry and the Ilse Katz Center for nanoscale Science and Technology, Ben-Gurion University of the Negev, Beer Sheva, Israel}

\begin{abstract}
Evaluating the time-dependent dynamics of driven open quantum systems is relevant for a theoretical description of many systems, including molecular junctions, quantum dots, cavity-QED experiments, cold atoms experiments and more. Here, we formulate a rigorous microscopic theory of an out-of-equilibrium open quantum system of non-interacting particles on a lattice weakly coupled bilinearly to multiple baths and driven by periodically varying thermodynamic parameters like temperature and chemical potential of the bath. The particles can be either bosonic or fermionic and the lattice can be of any dimension and geometry. Based on Redfield quantum master equation under Born-Markov approximation, we derive a  linear differential equation for equal time two-point correlation matrix, sometimes also called single-particle density matrix, from which various physical observables, for example, current, can be calculated. Various interesting physical effects, such as resonance, can be directly read-off from the equations. Thus, our theory is quite general gives quite transparent and easy-to-calculate results. We validate our theory by comparing with exact numerical simulations.  We apply our method to a generic open quantum system, namely a double-quantum dot coupled to leads with modulating chemical potentials. Two most important experimentally relevant insights from this are : (i) time-dependent measurements of current for symmetric  oscillating voltages (with zero instantaneous voltage bias) can point to the degree of asymmetry in the system-bath coupling, and (ii) under certain conditions, time-dependent currents can exceed time-averaged currents by several orders of magnitude, and can therefore be detected even when the average current is below the measurement threshold. 
\end{abstract}

\maketitle

\subsection{Introduction}
The dynamics of a quantum system coupled to an external environment (so-called ``open'' quantum system) are a result of the intricate interplay between the coherent evolution of the quantum degrees of freedom, the structure of the environment and the form and strength of the system-environment coupling. From quantum cavities \cite{ cavity1, cavity2, cavity3} and superconducting qubits \cite{supercond1,supercond2,supercond3,supercond4,supercond5,supercond6} , through quantum dots \cite{quantumdot1,quantumdot2,quantumdot3,quantumdot4,Gruss2016,Ajisaka2013}, molecular junctions \cite{Peskin2016, Peskin2010, Thingna2016, Segal2016, Ajisaka2015, Wang2015, Zelovich2015, Elenewski2017,Esposito2010,Hod2016, Rudge2016, Dubi2013} and  cold atoms \cite{coldatoms1,coldatoms2,coldatoms3,coldatoms4} , to excitons traveling in photosynthetic complexes \cite{Mohseni2014,Lambert2013,Fleming2011,Scholes2005}, open quantum systems show dynamics which can be far richer and more surprising than their coherent (environment-free) counterparts. 

Recent ideas of designing a quantum state by engineering a specific environment   \cite{KrausPRA2008,BarreiroNatPhys2010,Kienzler2015,Verstraete2009,
Diehl2008Nat,Bardyn2013NJP,Vorverg2013PRL} or  by a periodic modulation of the open system's Hamiltonian \cite{Volokitinarxiv2016,Hartmannarxiv2016,Peskin2016}  open a path to new forms of control over quantum systems. 
Combining these two concepts of time-periodic modulations and environment (bath) engineering, here we study the dynamics of open quantum systems where the environment thermodynamic parameters are periodically modulated. Even though there exists formally exact methods of treating such set-ups \cite{exact1,exact2,exact3,exact4,exact5,exact6,exact7,exact8,exact9}, they are generally quite difficult to use, bein computationally demanding. To simplify calculations, often the adiabatic approximation is used \cite{adiabatic1, adiabatic2, adiabatic3}. The adiabatic approximation assumes the environment parameters are modulated very slowly such that the system is always at a non-equilibrium steady state with the instantaneous environment. This approximation fails to capture dynamical effects which arise due to the time-periodic modulations. In this work, we therefore seek to go beyond the adiabatic approximation. 

For weak system-bath coupling, a combination of Quantum Master Equation methods (typically of the Lindblad form) with Floquet theory is often used to go beyond the adiabatic approximation \cite{floquet1,floquet2,floquet3,floquet4,floquet5,floquet6}. However, Floquet theory has the drawback of converting a finite-dimensional problem to an infinite dimensional one. Consequently, most interesting results from Floquet theory are often obtained as a perturbation expansion in terms of inverse drive frequency, giving results primarily for high frequency driving.

Moreover, from calculations in the absence of time-periodic modulations, the commonly used phenomenological Lindblad Quantum Master Equations are known to have several drawbacks, especially in an out-of-equilibrium situation \cite{Purkayastha2016,QME1,QME2,QME3}. It has been recently shown that microscopically derived Redfield Quantum Master Equation (RQME) under Born-Markov approximation can be used to overcome the drawbacks of phenomenological Lindblad equations and to obtain correct results up-to leading order in system-bath coupling \cite{Purkayastha2016}. To our knowledge, there has been no work in extending the RQME to the case when thermodynamic parameters of the baths are periodically modulated.

In this work, we derive a Born-Markov approximated Redfield theory for a quantum system consisting of non-interacting particles (bosons or fermions) on a lattice of arbitrary dimension and geometry, weakly coupled to multiple periodically-modulated baths. The baths are modeled by non-interacting particles (bosons or fermions) with infinite degrees of freedom. For a lattice of $N$ sites, this approach leads to a closed system of $N^2$ linear differential equations which can be easily solved numerically in time domain, thereby avoiding the drawback of infinite dimensional theory like Floquet theory.  We apply our formalism to a generic example of an open quantum system, namely a double quantum-dot coupled to two fermionic reservoirs, characterized by periodically modulated chemical potentials. The current through the system depends on both the chemical potential modulation properties and the chemical potential difference (i.e. voltage bias) between the electrodes. We show that even when there is no bias between the electrodes the dynamics of the system are non-trivial and reveal the internal structure of the junction (i.e. the internal spectrum of the dot and the asymmetry in the junction). When there is a voltage bias between the electrodes, the modulation induces unusual hysteresis behavior. These predictions can in principle be tested experimentally, demonstrating the applicability of our formalism to realistic open quantum systems.

\subsection{Set-up and protocol}\label{B}
Consider a general non-interacting tight-binding Hamiltonians for the whole system+bath set-up :

\begin{align}
\label{model_H}
\mathcal{H} &= \mathcal{H}_S + \mathcal{H}_B + \mathcal{H}_{SB}~,~~{\rm where} \\
~\mathcal{H}_S &= \sum_{\ell=1}^N H^{(s)}_{\ell m} \hat{a}_\ell^{\dagger} \hat{a}_m~,~~~
\mathcal{H}_B = \sum_{\ell=1}^N\sum_{r=1}^\infty \Omega_{\ell r} \hat{B}_{\ell r}^{\dagger} \hat{B}_{\ell r}~, \nonumber\\
\mathcal{H}_{SB} &= \varepsilon\sum_{\ell=1}^N \sum_{r} (\kappa_{\ell r} \hat{B}_{\ell r}^{\dagger} \hat{a}_\ell + \kappa_{\ell r}^* \hat{a}_\ell^{\dagger} \hat{B}_{\ell r})~.   
\end{align} 
$H^{(s)}$ is a Hermitian matrix and $\hat{a}_\ell$ correspond to fermionic (bosonic) annihilation operators defined respectively on $\ell$th lattice point of system and  $\hat{B}_{\ell r}$ to those of baths attached to the $\ell$th point.  
The baths  have infinite degrees of freedom. $\varepsilon$ is a dimensionless parameter that controls system bath coupling, so that $\{\kappa_{\ell r}\}$ have dimensions of energy. Here each lattice point is connected to a bath. By setting system-bath coupling to zero at an arbitrary number of lattice points, any geometry of system-bath connections can be achieved. We also introduce the bath spectral functions:
\begin{equation}
\label{J}
{J}_{\ell}(\omega)=2\pi \sum_r  \mid \kappa_{\ell r} \mid^2 \delta(\omega - \Omega_{\ell r}) 
\end{equation}

Note that the full system+bath Hamiltonian (\ref{model_H}) is time-independent. In the conventional dc bias case, for weak system-bath coupling, the baths are always assumed to be in thermal equilibrium with their own corresponding temperature and chemical potential. In other words, $Tr_B(\hat{B}_{\ell r}^{\dagger} \hat{B}_{\ell r})$, where $Tr_B(..)$ implies trace taken over bath degrees of freedom, is given by the corresponding fermi (bose) distribution $\Big [ e^{-\beta_\ell(\Omega_{\ell r}-\mu_\ell)}\pm1 \Big ]^{-1}$. Here, we are interested in the case when the inverse temperature $\beta_\ell$ and chemical potential $\mu_\ell$ are periodic functions of time. At this stage, one may want to model  such a set-up by deriving a quantum master equation for the density matrix of the system for the time independent case and making the fermi (bose) distributions time periodic `by hand'. However, the regime of validity of such an approach will not be clear. So, instead, in the following, we assume a protocol, and make physical assumptions so that such an equation may be reached. This shows how and under what conditions such a situation can arise physically.

We now describe the protocol. In the following, $\chi$ is the full density matrix of system+bath, $\rho_B$ is the density matrix of the bath, and $\rho$ is density matrix of the system obtained by tracing $\chi$ over bath variables. Let us, for the time being, assume one bath. The protocol can readily be generalized to multiple baths.

The protocol for a single bath is as follows :

a) At time $t=0$, $\chi = \rho \otimes \rho_B(0)$, with $\rho=\rho_0(0)$, which is some arbitrary initial system state, and $\rho_B(0) = {exp[-\beta(0)(\mathcal{H}_B-\mu(0)\mathcal{N})]}/{Z(\tau_S)}$. That is, the inital state is a product state of an arbitrary system state and a thermal bath state.

b) We evolve the system for a time $\tau_D$.  After time $\tau_D$, the system reaches the state $\rho_0(\tau_D)$. Note that during this time, the temperatures and chemical potentials of the bath has not changed.

c) At $t=\tau_D$, we restart the entire system+bath setup with the initial state $\chi = \rho \otimes \rho_B(\tau_D)$, with $\rho=\rho_1(0)=\rho_0(\tau_D)$ and $\rho_B(\tau_D) = {exp[-\beta(\tau_D)(\mathcal{H}_B-\mu(\tau_D)\mathcal{N})]}/{Z(\tau_D)}$. That is, at time $\tau_D$, the bath is changed into the thermal state with new inverse temperature and chemical potential $\beta(\tau_D), \mu(\tau_D)$, and the full system+bath state is again taken as the product state. This step implicitly assumes weak-system bath coupling. This is because, under weak system-bath coupling, we can assume that the bath is hardly affected by the system, and to leading order in system-bath coupling, the full density matrix is in product form.

d) We again let the system density matrix evolve under this new bath for time $\tau_D$ starting from the new initial state. Again, after time $\tau_D$, we restart the entire set-up with initial state $\chi = \rho(0) \otimes \rho_B(2\tau_D)$, with $\rho=\rho_2(0)=\rho_1(\tau_D)$ and $\rho_B(2\tau_D) = {exp[-\beta(2\tau_D)(\mathcal{H}_B-\mu(2\tau_D)\mathcal{N})]}/{Z(2\tau_D)}$.  
This protocol continues for a very long time and we will be mostly interested in the long time dynamics of this process.

In the limit of very small $\tau_D$, the above protocol gives a nearly continuous evolution of temperatures and chemical potentials of the baths. In particular, if the temperature and chemical potential vary periodically with a period $T$, and $\tau_D \ll T$, then the above protocol describes dynamics of a system evolving under a continuous periodic drive from the temperature and chemical potential of the bath. 

To make analytical progress with this problem, we need to make one more assumption, the Markov assumption. Let $\tau_B$ be the characteristic time-scale of relaxation of the effects of system-bath coupling on the bath. Then, we assume that $\tau_B \ll \tau_D$. It is only under this condition that microscopically derived Markovian Quantum Master Equations can be applied. So, if $\tau_{expt}$ be experimentally the smallest time scale, most of our following discussion holds  if the following condition on time-scales is maintained :
\begin{align}
\label{t_cond1}
\tau_B \ll \tau_D \ll \tau_{expt} \ll T
\end{align}
It can be shown that $\tau_B$ depends on the temperature of the bath, $\tau_B \sim \beta$ (see Appendix~\ref{Appendix_B}). Hence, our analytical discussion below will not be valid at extremely low temperatures. It is only valid when temperature is large enough so that above condition on time scales can be satisfied.

For multiple baths, the above protocol is followed for each bath. $\tau_B$ is then taken as the largest of the relaxation times of the baths, so that $\tau_D$ is much larger than relaxation time scales of all baths. When the system is driven by multiple baths according to our protocol, we call it an ac drive process. Correspondingly, the usual case of having a time independent temperature or chemical potential difference between baths is called a dc bias process.

The above protocol, along with the condition (Eq.~\ref{t_cond1}) on time scales, breaks down the ac drive process into steps of time independent processes where Born-Markov approximation can be applied. For such time independent processes, the Born-Makov approximated Redfield Qunatum Master Equation (RQME), as well as the evolution equation for two point correlation functions from the RQME have been derived for Hamiltonian (\ref{model_H}) recently \cite{Purkayastha2016} (see Appendix~\ref{Appendix_A}). Going through the above protocol with the  RQME only has the effect of making the fermi (bose) distribution functions time-dependent with the instantaneous temperatures and chemical potentials (see Appendix~\ref{Appendix_C}). The final result, as desired, will be same as that obtained by `putting by hand' time dependent fermi of bose distribution functions in the RQME derived without such time dependence. However, it is important to note that, the description via such an equation would not be possible if each step of our protocol would not satisfy the Markov condition $\tau_B \ll \tau_D$. Hence, only when system-bath coupling is weak and the condition on time scales in Eq.~\ref{t_cond1} is satisfied, can our set-up and protocol be described by such an equation. This crucial fact would not be clear in a `putting by hand' approach. Also, for small systems, the protocol can easily be exactly (i.e, without Markov approx) simulated with finite but large baths, and hence allows for numerical validation of results. 

\subsection{Redfield equation for the correlation matrix (single particle density matrix)}\label{C}

Since the system is quadratic, various physical observables like current and occupation can be directly calculated from the two-point correlation functions. Using the results in Ref.~\cite{Peskin2016} (which are also re-derived in Appendix),  we can readily write down the evolution equation for two-point correlation functions for our set-up. For this, it is convenient to go to the eigenbasis of the system Hamiltonian. Let $c$ be the unitary matrix which diagonalizes $H^{(S)}$, i.e., \begin{equation}
c^\dagger H^{(S)}  c= \omega^{(D)}~,
\end{equation}
where  $c^\dagger c=I$ and $\omega^{(D)}$ is a diagonal matrix with elements $\omega_\nu$.  Then we also define new operators $\{A_\alpha\}$ through the transformation \begin{equation}
\hat{a}_{\ell} = \sum_{\alpha=1}^N c_{\ell \alpha} \hat{A}_\alpha~.
\label{Aop}
\end{equation}
 Thus $\hat{A}_\alpha$ is the annihilation operator for the $\alpha$th eigen-mode with energy $\omega_\alpha$.  

The evolution equation for the equal time two point correlation functions $C_{\alpha\nu}(t) = \langle \hat{A}_\alpha^\dagger(t) \hat{A}_\nu(t) \rangle = Tr\big( \rho\hat{A}_\alpha^\dagger(t) \hat{A}_\nu(t) \big)  $ can be derived by following the derivation of the Redfield equation for the density matrix $\rho$ \cite{breuer2007theory,CarmichaelBook,Nitzan2006} and substituting it back to the definition of the correlation functions \cite{Purkayastha2016}(see Appendix~\ref{Appendix_A}).  The protocol, along with Born-Markov approximation, only makes the fermi (bose) distributions time-dependent (see Appendix~\ref{Appendix_B}). The resulting equation is
\begin{align}
\label{C_differential}
&\frac{dC_{\alpha\nu}}{dt} = \Big [ i\omega_\alpha C_{\alpha\nu}(t)  -\varepsilon^2  \sum_{\sigma=1}^N C_{\alpha\sigma}(t) v_{\nu \sigma }+(\alpha \leftrightarrow \nu )^\dagger \Big ]\nonumber \\
&+ \varepsilon^2 Q_{\alpha\nu}(t)~,  \\
&~~{\rm where}, \nonumber \\ 
& v_{\alpha \nu}=f_{\alpha\nu}(\omega_{\nu})-if_{\alpha\nu}^{\Delta}(\omega_{\nu}) \nonumber \\
& Q_{\alpha\nu}(t) =  [F_{\nu\alpha}(\omega_\alpha,t)-iF_{\nu\alpha}^{\Delta}(\omega_\alpha,t)+(\alpha \leftrightarrow \nu )^*] \nonumber \\
&f_{\alpha\nu}(\omega) = \sum_{{\ell}=1}^Nc_{\ell \alpha}^*c_{\ell \nu} \frac{J_{\ell}(\omega)}{2}, \nonumber \\
& F_{\alpha\nu}(\omega,t) = \sum_{{\ell}=1}^N c_{\ell \alpha}^*c_{\ell \nu} \frac{J_{\ell}(\omega)n_{\ell}(\omega,t)}{2} \nonumber
\end{align}
with $f_{\alpha\nu}^{\Delta}(\omega) = \mathcal{P} \int \frac{}{}\frac{d\omega^{\prime} f_{\alpha\nu}(\omega^{\prime})}{\pi(\omega^{\prime}-\omega)}, \hspace{2pt}
F_{\alpha\nu}^{\Delta}(\omega,t) = ~ \mathcal{P} \int \frac{}{}\frac{d\omega^{\prime} F_{\alpha\nu}(\omega^{\prime},t)}{\pi(\omega^{\prime}-\omega)}$, where $\mathcal{P}$ denotes principal value, $n_\ell(\omega,t)~=~\Big [ e^{-\beta_\ell(t)(\omega-\mu_\ell(t))}\pm1 \Big ]^{-1}$ is the instantaneous fermi or bose distribution function and $J_\ell(\omega)$ is the spectral function (Eq.~\ref{J}) of bath (lead) attached to $\ell$th site of the system. Weak system-bath coupling requires $\varepsilon^2 J_{\ell}(\omega) \ll \{\omega_{\alpha} \}$.

Eq.~\ref{C_differential} gives a closed set of $N^2$ linear differential equations. The matrix with elements $C_{\alpha\nu}$ is sometimes also called the single particle density matrix of the system. Writing $C$ and $Q$ as vectors, this equation can be cast in the form 
\begin{align}
\label{C_vector_differential}
&\frac{dC}{dt}=-MC+ \varepsilon^2 Q(t),\\
&{\rm where} \nonumber \\
& M = \mathbb{I}_N\otimes G^* + G \otimes \mathbb{I}_N, \nonumber \\
& G_{\alpha\nu} = \varepsilon^2 v_{\alpha\nu}^* - i\omega_{\nu}\delta_{\alpha\nu} \nonumber 
\end{align}
$\mathbb{I}_N$ is the $N$ dimensional identity matrix and $\otimes$ denotes Kronecker product. Eq.~\ref{C_vector_differential} has the formal solution
\begin{align}
\label{C_vec_t}
C(t) = e^{-Mt}C(0) + \varepsilon^2\int_0^t dt^\prime e^{-M(t-t^\prime)}Q(t^\prime) 
\end{align}

Note that the matrix $M$ has no time dependence and hence is the same matrix as would appear in the dc problem. If under dc bias the system reaches a steady state after a long time, the real part of eigenvalues of the matrix $M$ has to be positive. We will assume this henceforth.

We wish to look at the long time properties of Eq.~\ref{C_vec_t}.
Let our ac drive be periodic with a time period $T$. Then, for integer $r$, 
\begin{align}
Q(t+rT) = Q(t), \hspace{5pt} r\in \mathbb{Z}
\end{align}
We break up the observation time $t$ into steps of $T$, so that, for integer $m$,
\begin{align}
t=mT+\tau \hspace{5pt} m\in \mathbb{Z}
\end{align}
Then Eq.~\ref{C_vec_t} can be written as
\begin{align}
&C(mT+\tau) = \varepsilon^2\sum_{r=1}^m \Big ( e^{-MrT} \Big ) \int_0^T dt^\prime e^{-M(\tau-t^\prime)}Q(t^\prime) \nonumber \\
&+\varepsilon^2\int_0^\tau dt^\prime e^{-M(\tau-t^\prime)}Q(t^\prime)+e^{-M(mT+\tau)}C(0)
\end{align}
Now, since we have assumed that the real part of all the eigenvalues of $M$ are positive, we can perform the sum in above equation. Also, we are interested in the long time limit,  $m \gg 1$. Hence we have
\begin{align}
\label{C_vec_steady}
&C(mT+\tau) = \varepsilon^2(e^{MT}-\mathbb{I}_{N^2})^{-1} \int_0^T dt^\prime e^{-M(\tau-t^\prime)}Q(t^\prime) \nonumber \\
&+\varepsilon^2\int_0^\tau dt^\prime e^{-M(\tau-t^\prime)}Q(t^\prime)
\end{align} 
Note that, in the long time limit, the RHS is independent of $m$, and is also independent of the initial condition. This means that in the long time limit, the correlation functions settle down in a periodic state, with period same as the ac drive. This is consistent with Floquet theory. However, note that, Floquet theory was not used explicitly to derive this result. Eq.~\ref{C_vec_steady} is the central result of the manuscript. 

\subsection{Resonances and Currents}\label{D}

Now, let us go back to Eq.~\ref{C_differential}. Since in the long time limit the periods of $C(t)$ and $Q(t)$ are same, we can perform a Fourier series expansion :
\begin{align}
C_{\alpha\nu}(t) = \sum_{p=-\infty}^\infty C_{\alpha\nu}^p e^{ip\Omega_0 t}, \hspace{2pt} Q_{\alpha\nu}(t) = \sum_{p=-\infty}^\infty Q_{\alpha\nu}^p e^{ip\Omega_0 t} 
\end{align}
where $\Omega_0=\frac{2\pi}{T}$. Substituting in Eq.~\ref{C_differential}, we obtain the following equation for each Fourier mode,
\begin{align}
\label{C_Fourier}
&i(\omega_\alpha -\omega_\nu-p\Omega_0) C_{\alpha\nu}^p + \varepsilon^2 Q_{\alpha\nu}^p \nonumber \\
& -\varepsilon^2 \Big [ \sum_{\sigma=1}^N C_{\alpha\sigma}^p v_{\nu \sigma }+C_{\sigma\nu}^p v_{\alpha\sigma }^*~ \Big ] =0
\end{align}

This immediately gives us two important results. First, we note that for the time-average correlation functions, $\overline{C}_{\alpha\nu}(t) = \frac{1}{T}\int_t^{t+T} dt^\prime C_{\alpha\nu}(t^\prime)=C_{\alpha\nu}^0$, and similarly for $Q_{\alpha\nu}$. With this we find that the equation averaged over one time period gives exactly the steady state equation for a dc bias given by the time-period averaged bose or fermi distribution functions $\overline{n_\ell(\omega,t)}$.

Second, we see that when $\omega_\alpha -\omega_\nu=p\Omega_0$, Eq.~\ref{C_Fourier} becomes independent of system-bath coupling $\varepsilon^2$. This is the phenomenon of resonance. At resonance, the leading term of system correlation functions do not depend on system bath coupling. So all system properties, like current between two sites inside the system, which were otherwise proportional to $\varepsilon^2$ in the leading term, become larger by orders of magnitude.

Even though system properties become large at resonance, the current from any of the baths still remain small. To see this, we need the expression for current from the baths in terms of the correlation functions. The expression for particle current is obtained from the continuity equation for conservation of particles
\begin{align}
\label{current_B}
&\frac{d}{dt}\Big (\sum_{\alpha=1}^N C_{\alpha\alpha} \Big ) = \sum_{\ell} I_{B^{(\ell)}\rightarrow \ell}
\end{align} 
$I_{B^{(\ell)}\rightarrow \ell}$  is the particle current from bath attached to the $\ell$th site. Under dc bias, in steady state, the LHS of above equation is zero, and hence the currents from the baths are equal. However, under ac drive, even at long time, the LHS is not zero, and thus, instantaneous currents from the baths can be different.  The summation on the right runs over all lattice sites connected to bath. The expressions for the currents from the baths obtained from above continuity equations and Eq.~\ref{C_differential} are
\begin{align}
& I_{B^{(\ell)}\rightarrow \ell} = \varepsilon^2 \Big [ \sum_{\alpha=1}^N Q_{\alpha}^{(\ell)}(t) - \sum_{\sigma,\alpha=1}^N C_{\alpha\sigma}(t)\big(v_{\alpha\sigma}^{(\ell)} + v_{\sigma\alpha}^{(\ell)*}\big) \Big ]
\end{align}

where 
\begin{align}
&Q_{\alpha}^{(\ell)}(t) = |c_{\alpha \ell}|^2  \mathfrak{J}_\ell(\omega_\alpha)n_\ell(\omega_\alpha) \nonumber \\
&v_{\alpha\sigma}^{(\ell)} = c_{\alpha \ell}^* c_{\sigma \ell} \Big(\frac{\mathfrak{J}_\ell(\omega_\sigma)}{2}- i \mathcal{P} \int \frac{d\omega}{2\pi}\frac{\mathfrak{J}_\ell(\omega)}{\omega-\omega_\sigma} \Big )
\end{align}

These expressions show that the currents from the baths are explicitly proportional to $\varepsilon^2$. So even when the system correlation functions are independent of $\varepsilon^2$ in the leading order, the currents from the baths are still $O(\varepsilon^2)$. However, this is not true for particle current between two sites which reside \emph{within} the system. 
The current between $\ell$th and $\ell+1$th lattice sites of the system is given by
\begin{align}
\label{current_S}
I_{\ell \rightarrow \ell+1} = 2 \hspace{2pt} Im \Big ( H^{(s)}_{\ell \hspace{2pt} \ell+1} \hat{a}_\ell^{\dagger} \hat{a}_{\ell+1} \Big )
\end{align}  

This current is not explicitly proportional to $\varepsilon^2$. At resonance, it is independent of $\varepsilon^2$ in the leading order. Thus, at resonance, the current between two adjacent lattice points in the system can be much larger than the current from the baths. On the other hand, the time-period averaged current, which corresponds to steady state of a dc bias, is same inside the system as from the baths.

The difference in frequency between adjacent resonances, which is given by
\begin{align}
\frac{\omega_\alpha-\omega_\nu}{p} -\frac{\omega_\alpha-\omega_\nu}{p+1} = \frac{\omega_\alpha-\omega_\nu}{p(p+1)} \sim \frac{1}{p^2}
\end{align}
decreases as $1/p^2$. Hence small driving frequencies are always close to resonance with one of the higher modes (large $p$) of the steady state oscillations. So, for small driving frequencies, there is not much difference between the resonance and off-resonance condition. The values of the correlation functions, however, may not be so large as the first resonance ($p=1$). This is because, the weight of the driving signal at higher modes may decrease. 

It is also interesting to note that none of the expressions for currents can be reduced to the form of difference between fermi (bose) distributions of the various baths. Therefore, even when all the baths are driven by the exactly same time dependent  temperature or chemical potential (symmetric ac drive), so that there is no instantaneous temperature or chemical potential difference, there can be an instantaneous current, both between the system and the baths and inside the system. At resonance, the internal currents may have a large amplitude (compared to the system-bath currents). The time-period averaged current is, of course, zero in this case.

It is important to state that we have \textit{not} made the ubiquitous adiabatic approximation $T \gg t_{steady}$, where $t_{steady}$ is the time to reach steady state, which corresponds to the smallest real part of eigenvalues of the matrix $M$ in Eq.~\ref{C_vector_differential}. If adiabatic approximation were made, then the expressions for the correlation functions at any time would be given by the dc-bias steady state results with the fermi (bose) distributions given by the instantaneous temperatures and chemical potentials. In that case, the expressions for currents \textit{would have} reduced to the form of difference between fermi (bose) distributions of the various baths, and \textit{no instantaneous current} would have been seen in the symmetric ac drive case.

Under dc bias, the steady state can be quite easily obtained for non-interacting open systems described by Hamiltonian of the form Eq.~\ref{model_H} by exact methods (i.e, not under Born-Markov approximation). Transient dynamics of approach to steady state, on the other hand, is generally quite difficult to calculate from such methods. From our protocol, it can be seen that when,  $T \lesssim t_{steady}$ (i.e, when adiabatic approximation is not valid), it is the transient behaviour of approach to steady state for dc bias that becomes important to describe the ac driven case. Therefore, such exact methods are difficult to use in our ac drive set-up.   However, as has been recently established, RQME gives the correct time dynamics of approach to steady state under dc bias of two-point correlation functions for our set-up as long as Born-Markov approximation is satisfied \cite{Purkayastha2016}. This fact has allowed us to use RQME for ac drive case to go beyond the adiabatic approximation easily.

Recently, another method based on a quantum master equation called the Discrete Lioville von-Neumann (DLvN) method has been formulated \cite{DLvN1,DLvN2,DLvN3}, which has been shown to give accurate results for time-dependent behaviour \cite{DLvN3}. However, in this method one models the leads via finite but large size Hamiltonians, and needs to solve explicitly for the leads along with the system. As a result, if there are $N_L$ sites in left lead, $N_R$ sites in the right lead and $N$ sites in the system, then one needs to solve for a closed set of $(N_L + N_R + N)^2$ linear differential equations. In comparison, using our RQME method, the leads are considered implicitly in the spectral functions (Eq.~\ref{J}) and one needs to solve only for a closed set of $N^2$ linear differential equations (see Eqs.~\ref{C_differential} and \ref{C_vector_differential}). 
  
RQME has the drawback of not being completely positive because of not being of the Linblad form. As a result, under certain initial conditions, it may lead to a non-physical state at short times.  In fact, the rigorous microscopic derivation of the dc bias Redfield equation for quadratic baths requires the observation time $t$ to satisfy $t\gg \tau_B$  (see Appendix~\ref{Appendix_A}). Thus, it is only in this regime of observation times that the results should be valid. As a consequence, one expects that, problems regarding non-positivity show up only at short times ($t\leq \tau_B$), where the Redfield equation itself is not justified \cite{Eastham2016}. Based on this, it is expected that, in our ac drive set-up, when the condition on time scales in Eq.~\ref{t_cond1} is satisfied, problems regarding positivity will not arise. This, however, is not a completely rigorous statement, and requires further investigation. In Ref \cite{Purkayastha2016}, the long time behaviour of the correlation functions for the dc bias case of our set-up have been shown to be given correctly by the Redfield equation. Since our protocol breaks down the ac drive process into a series of such dc drive steps, this gives further support regarding validity of our results. But, as mentioned before, since $\tau_B$ is inversely proportional to temperature (see Appendix~\ref{Appendix_B}), our theory may not be valid at extremely low temperatures.

\subsection{A detailed example: the $N=2$ case}\label{E}
  
All the above results are quite general and hold for non-interacting system of bosons or fermions in lattice of any dimension and geometry.  In the following, to validate the theory, as well as to better understand the physics of such ac drive, we apply this theory to a simple model of two fermionic lattice sites, each connected to their own baths. This generic model can be used to describe physical systems such as driven double quantum dots\cite{Stehlik2016} or single-molecule junctions with two molecular moieties (e.g. biphenyl-dithiol molecular junctions \cite{Venkataraman2006}). 

We consider the following specific two-site system  coupled to baths which are one-dimensional chains:
\begin{align}
\hat{\mathcal{H}}_S &= \omega_{0} (\hat{a}_1^{\dagger} \hat{a}_1+\hat{a}_2^{\dagger} \hat{a}_2) + g(\hat{a}_1^{\dagger} \hat{a}_{2} + \hat{a}_{2}^{\dagger} \hat{a}_1)~, \nonumber \\
\hat{\mathcal{H}}_B^{(\ell)}&= t_B (\sum_{s=1}^\infty \hat{b}_s^{\ell\dagger}\hat{b}_{s+1}^\ell+h.c.), \hspace{5pt} \hat{\mathcal{H}}_B = \hat{\mathcal{H}}_B^{(1)}+\hat{\mathcal{H}}_B^{(2)}~,\nonumber \\
\hat{\mathcal{H}}_{SB} &= \varepsilon\gamma_1 (\hat{a}_1^{\dagger}\hat{b}_{1}^1+ h.c.) + \varepsilon\gamma_2 (\hat{a}_2^{\dagger}\hat{b}_{1}^2+ h.c.)~,  \label{ham2S} 
\end{align}
where the operators are all fermionic and $\hat{b}_{s}^\ell$ is the annihilation operator of the $s$th bath site of the $\ell$th bath. The eigenmodes of the system are given by $\hat{A}_1 = {(\hat{a}_1-\hat{a}_2)}/{\sqrt{2}}$, $\hat{A}_2 = {(\hat{a}_1+\hat{a}_2)}/{\sqrt{2}}$ with eigenvalues $\omega_1 = \omega_0 - g$, $\omega_2 = \omega_0 + g$. The bath spectral functions, defined in Eq.~\ref{J}, can be obtained explicitly by going to eigenmodes of the baths and are given by \cite{Purkayastha2016,Peskin2016} (see Appendix~\ref{Appendix_D})
\begin{equation}
\mathfrak{J}_\ell(\omega) = \Gamma_\ell\sqrt{1-\left(\frac{\omega}{2t_B}\right)^2}~, \hspace{5pt} \Gamma_\ell=\frac{2\gamma_\ell^2}{t_B}~.
\label{Jform}
\end{equation}
$\omega_0 \gg (\varepsilon^2/t_B)$ so that QME can be applied while the parameter $g$ can be varied freely. The two baths are taken at the same temperature. Inverse temperature $\beta$ is taken to be constant and the ac drive is given by periodic chemical potential :
\begin{align}
n_\ell(\omega,t) = [e^{\beta(\omega-\mu_\ell(t))}+ 1]^{-1}, \hspace{5pt} \mu_\ell(t+rT)=\mu_\ell(t) \hspace{2pt} \forall r\in \mathbb{Z}
\end{align}

We choose the drive frequency as 
\begin{align}
\label{p_def}
\Omega_0 = \frac{\omega_2 - \omega_1}{p} = \frac{2g}{p}
\end{align}

If $p$ is integer, the system is at one of the resonances. If $p$ is not an integer, it is away from resonance.  

We look at the particle current in the ac driven steady state. The expressions for the currents are given in Eq.~\ref{current_B},~\ref{current_S}. For the $N=2$ case, the expression for $I_{1\rightarrow 2}$ simplifies to $I_{1\rightarrow 2}= 2g\hspace{2pt} Im (C_{12})$.
Two different ac drives are considered: 

(i) symmetric ac drive or zero voltage drive :
\begin{align}
\label{pot_num_0}
 \mu_1 = \mu_2 = V_0 \sin(\Omega_0 t)
\end{align}

(ii) asymmetric ac drive :
\begin{align}
\label{pot_num_1}
\mu_1 = V_1 \cos(\Omega_0 t), \hspace{5pt} \mu_2 = V_2 \sin(\Omega_0 t)
\end{align}

\begin{figure}[t]
\includegraphics[width=9cm,height=9cm]{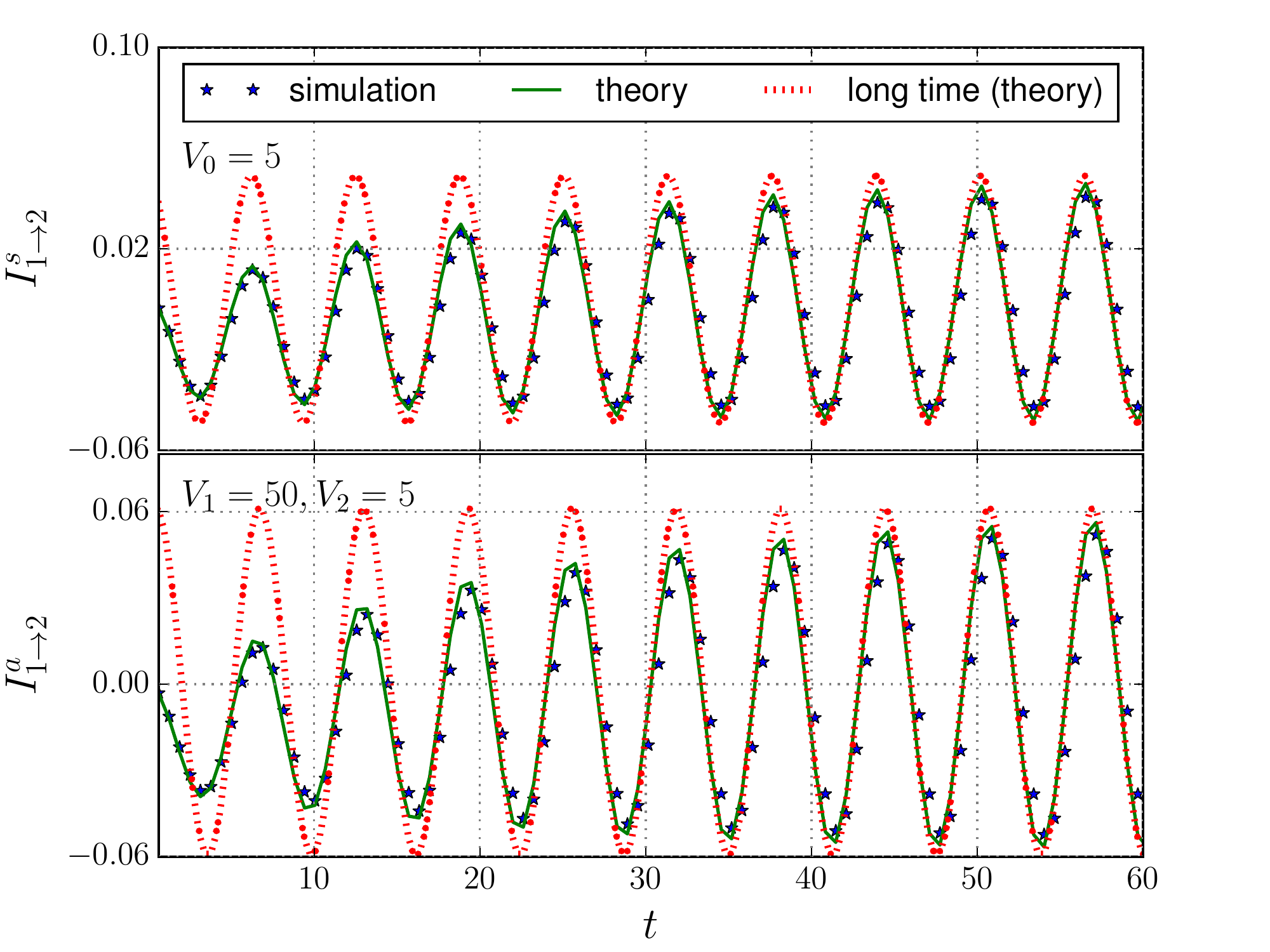} 
\caption{Internal current $I_{1\rightarrow_2}$ as a function of time, for the two cases (i) for symmetric ac drive: $\mu_1=\mu_2=V_0 \cos(\Omega_0 t)$ (top panel), (ii) for asymmetric ac drive :$\mu_1 = V_1 \cos(\Omega_0 t), \hspace{5pt} \mu_2 = V_2 \sin(\Omega_0 t)$ (bottom panel). The figure compares numerically exact simulation results for current inside the system (stars) with those obtained from our theory Eq.~\ref{C_vec_t} for all times (solid line), and Eq.~\ref{C_vec_steady} for long times (dashed line). The near perfect agreement validates our theory.  Other parameters $\Omega_0=2g, g=0.5,t_B=200,\omega_0=1,\beta_1=\beta_2=0.1,\Gamma_1=0.01,\Gamma_2=0.09$. All times are measured in units of $\omega_0^{-1}$, all energies are measured in units of $\hbar \omega_0$.}
 \label{fig:one_simu}
\end{figure}

\begin{figure*}[t]
\includegraphics[scale=0.55]{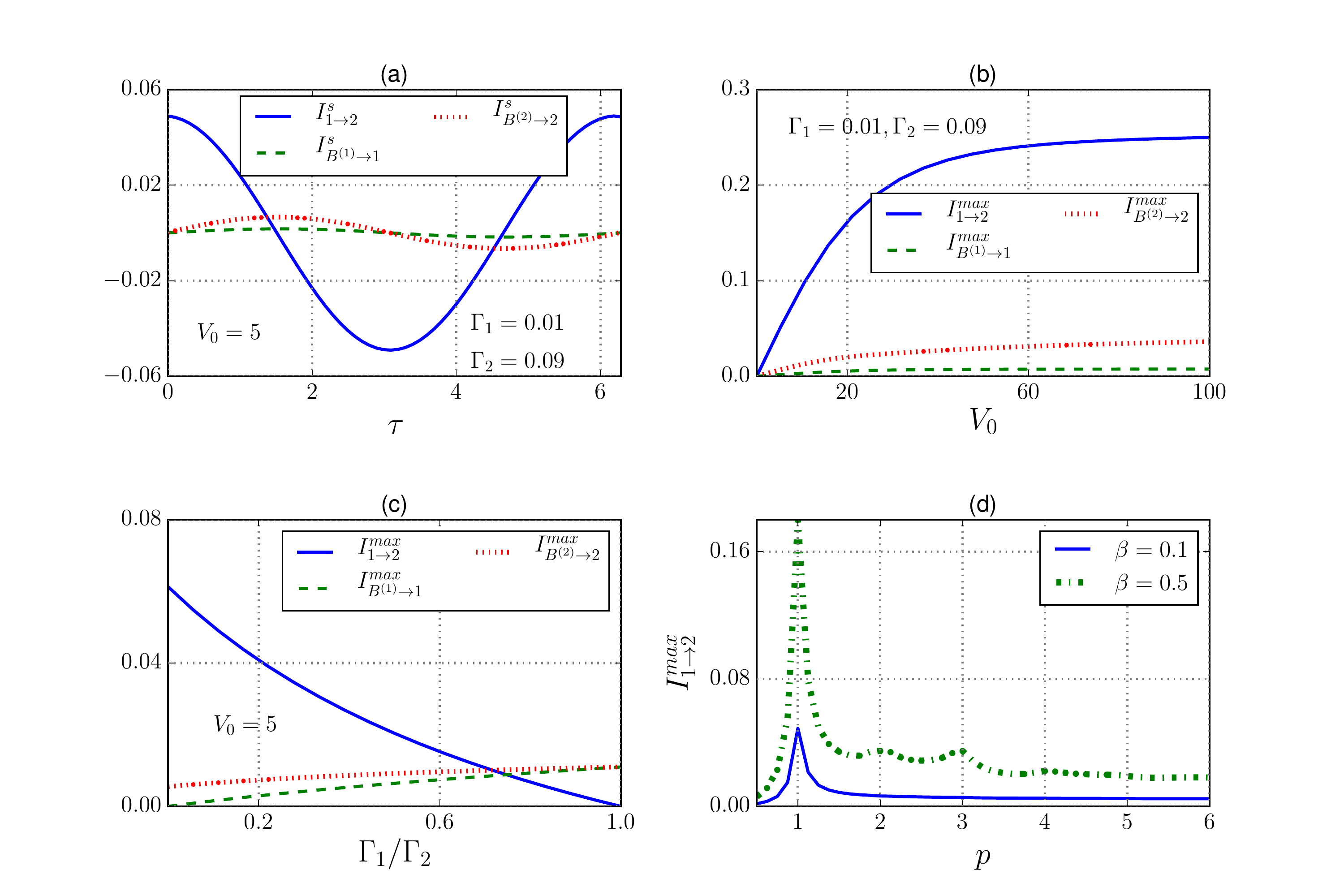} 
\caption{Various currents under  symmetric ac drive ($\mu_1~=~\mu_2~=~V_0\sin(\Omega_0 t)$). Panel (a), (b), (c) describes the set-up at first resonance ($\Omega_0=\omega_2-\omega_1=2g$).  Panel (a) shows long time results of $I_{1\rightarrow 2}^{s}$ and currents from the baths ($I_{B^{(1)} \rightarrow 1}^{s}$,$I_{B^{(2)} \rightarrow 2}^{s}$) with asymmetric system-bath coupling ($\Gamma_1 \neq \Gamma_2$) as a function of time over one time period of the ac drive. $I_{B^{(1)} \rightarrow 1}^{s}$,$I_{B^{(2)} \rightarrow 2}^{s}$ are much smaller than $I_{1\rightarrow 2}^{s}$.  Panel (b) shows the behavior of the maximum currents with $V_0$ for asymmetric system-bath coupling. The maximum currents increase with $V_0$ and finally saturates. Panel (c) shows behavior of the maximum currents with the degree of asymmetry ($\Gamma_1/\Gamma_2$). The maximum current inside the system ($I_{1\rightarrow 2}^{max}$) decreases with increase in degree of asymmetry, and becomes zero for symmetric coupling. In contrast, maximum current from baths ($I_{B^{(1)} \rightarrow 1}^{max}$,$I_{B^{(2)} \rightarrow 2}^{max}$) remain non-zero and become same for symmetric coupling. This shows a stark difference between symmetric and asymmetric system-bath coupling, and gives an experimental way to determine asymmetry of system-bath coupling. Panel (d) shows the maximum current in the system as a function of $p=2g/\Omega_0$ (Eq.~\ref{p_def}) at two different temperatures. $p$ being integer corresponds to resonances. The first resonance peak is very strong. Higher resonance peaks are much weaker, and are washed out by increasing temperature.   Other parameters $g=0.5,t_B=200,\omega_0=1,\beta_1=\beta_2=0.1$. All times are measured in units of $\omega_0^{-1}$, all energies are measured in units of $\hbar \omega_0$.   } \label{fig:sym_drive}
\end{figure*}

\subsubsection{Comparison between analytic formula and exact numerics}
Before proceeding to elucidate the physics dominating the driven system, we wish to validate the analytic derivation. For this we compare results obtained from Eq.~\ref{C_vec_t} with that obtained from full numerical simulation of our protocol. For small systems like the two-site case considered here, the protocol for our ac drive set-up can be simulated exactly with finite but large baths. For each time-independent step of our protocol, we choose a bath of finite size with bath correlations satisfying fermi  distributions and evolve the full system+bath Hamiltonian $\hat{\mathcal{H}}$ using unitary Hamiltonian dynamics. Let us collectively denote by ``$d$'' a column vector with all annihilation operators of both system and baths. The full Hamiltonian can be written as $\hat{\mathcal{H}}=\sum_{i,j}H_{i j} d^\dagger_i d_j$ where $i$ now refers to either system or bath sites. If  $ D= \langle d d^{\dagger} \rangle$ denotes the full correlation matrix of system and baths, its time evolution is given by $D(t)=e^{i H t} D e^{-i H t}$. The bath correlations are then changed according to the protocol, and the process is repeated. Various observables like current inside the system calculated using this exact numerical simulation can be compared with that obtained from Eq.~\ref{C_vec_t}, thus providing a way to validate our theory. Note that the numerical simulation does not take into consideration the `Markov' condition on time scales given in Eq.~\ref{t_cond1}. It holds even when Eq.~\ref{t_cond1} is not respected. Thus it allows us to check the validity of the crucial assumption on time scales required for our analytical treatment.

Fig.~\ref{fig:one_simu} shows the results for current inside the system $I_{1\rightarrow 2}^{s}$ and $I_{1\rightarrow 2}^{a}$ for cases (i) Eq.~\ref{pot_num_0} and (ii) Eq.~\ref{pot_num_1} respectively (superscripts in $I$ stand for symmetric and asymmetric currents), as obtained from exact numerics as well as from our theory. Eq.~\ref{C_vec_steady}  has been used to obtain the long time result, while Eq.~\ref{C_vec_t} has been used to get the result at all times, showing approach to the long time dynamics. Numerical simulation has been done with baths of size $256$ sites, which are large enough to have negligible finite-size effects.  The near perfect agreement with exact numerical simulations validate our theory. For this plot, the drive frequency is chosen to be $\Omega_0=\omega_1-\omega_2=2g$, so that the set-up is at the first resonance.  The near perfect match occurs for other frequencies also as long as Eq.~\ref{t_cond1} is satisfied. For our choice of parameters $\beta\sim 0.1$ satisfies Eq.~\ref{t_cond1}. The initial condition for plots shown in the figure corresponds to no particle in the system, but the agreement with numerical simulations has been checked for other initial conditions (like randomly chosen initial values of correlation functions etc.) also.

Note that numerical validation was only possible owing to the small size of the system, which allowed for using reasonably sized finite baths. For larger system sizes, much larger baths will be required and the set-up will not be amenable to numerical simulation. However, the theory can be easily used for much larger systems connected to infinite baths. 

Having validated the theory, we now look at the physics of the long time dynamics for both the cases.

\subsubsection{Symmetric - or zero voltage - ac drive}

First, we look at long time dynamics of the symmetric drive (Eq.~\ref{pot_num_0}). In this case, both chemical potentials are the exact same sinusoidal so that there is no instantaneous voltage difference, i.e, $V(t)=\mu_1(t)-\mu_2(t)=0$. Even though there is no voltage difference, as discussed before, because of being a driven system, there can still be an instantaneous current. Moreover, because of presence of displacement current, the current from the left bath, current in the system, current from the right bath can be different. 

Panel (a) of Fig.~\ref{fig:sym_drive} shows currents from the left (right) bath to site 1(2), $I^s_{B^{(1)}\rightarrow 2}~(I^s_{B^{(2)}\rightarrow 2})$,  and current inside the system over one time period for the case of first resonance ($p=1$ in Eq.~\ref{p_def}). As expected from our discussion of resonance, the current inside the system is much larger than the currents from the baths. A physical explanation for the non-zero instantaneous current for zero voltage drive can be given as follows.

Let $V_0>\omega_2$ (see Eq.~\ref{pot_num_0}). As the chemical potential varies, the Fermi energy of the particles in the bath varies. When it exceeds $\omega_2$, particles flow into the system from the baths. When the Fermi energy of the particles in the bath is smaller than $\omega_1$, particles flow out of the system into the baths. If the time to reach steady state is larger than the time period of the drive, this transient behavior is observed, which leads to the instantaneous current. 

It is clear from this argument, that the instantaneous current increases with increase in $V_0$. Particularly, if $V_0<\omega_1$, there will  be a small instantaneous current. This behaviour is shown in panel (b) of Fig.~\ref{fig:sym_drive} which shows the maximum instantaneous currents from the baths and inside the system as a function of $V_0$. The current increases continuously with $V_0$ and finally saturates. The saturation occurs because there are only two eigen-energy levels of the system. Actually, because of the fermionic nature of the set-up, one would expect steps or kinks at the positions  $V_0=\omega_1$ and $V_0=\omega_2$. However, such behaviour is not observed because temperature is not low enough.  On the other hand, it is important to note that the time-period averaged current is zero always because it is proportional to the difference between time-period averaged fermi distributions of the two baths.  

We note that the internal current, while strictly speaking cannot be measured directly, nevertheless may have physical consequences. Specifically, such internal current may lead to local heating of the junction (due to Joule heating). This can lead for instance, to a breakdown of the system (in molecular junctions) or to heating-induced observable changes in current (for a double quantum dot).

Panel (c) of Fig.~\ref{fig:sym_drive} shows a more interesting effect. The plot shows the maximum instantaneous current from the baths and inside the system as a function of asymmetry $\Gamma_1/\Gamma_2$ of the system-bath coupling (note that the chemical potentials are still symmetric). The set-up is still at first resonance. The maximum current inside the system decreases with decrease in asymmetry, and vanishes for the fully symmetric junction, $\Gamma_1/\Gamma_2=1$. At the same time, in the symmetric point the maximum currents from left and right baths are equal. This can be physically explained as follows. If the baths are identically coupled as well as identically driven, the particles come into the system at exactly same rate leading to same current from both baths. Since the particles are fermions, if one particle occupy each site, there can be no current in between the two sites due to Pauli exclusion principle. This leads to the fact that if rate of inflow of particles from both baths is same, there is no current between the two sites. Thus, current between the two sites of the system  comes from a mismatch between rate of inflow of particles from the baths. Therefore, current inside the system increases with increase in asymmetry of system-bath coupling. 

Panel (d) of Fig.~\ref{fig:sym_drive} shows plots of the maximum current inside the system for asymmetric system-bath coupling with $p$ (as defined in Eq.~\ref{p_def}) for two different temperatures. When $p$ is equal to an integer, the set-up is at resonance. Thus, from our previous discussion, the maximum current inside the system should show peaks at integer values of $p$. The first resonance peak is strong. However, lower frequency resonance peaks are much weaker. This is because the driving signal itself has a small contribution from such modes. Moreover, we see that higher temperature washes out the lower frequency resonances. 

We point out that asymmetry in molecular junctions may be a key ingredient in deciphering its electronic transport properties and in designing single-molecule  devices (see, e.g., \cite{Warner2015,Zhang2015,Wang2014}). Our results demonstrate that measuring the time-dependent current (for zero voltage bias but time-dependent voltages) is a direct way to measure asymmetry in molecular junctions, which can serve (in parallel to usual transport measurements) for junction characterization.

\begin{figure}[t]
\includegraphics[width=9cm,height=8cm]{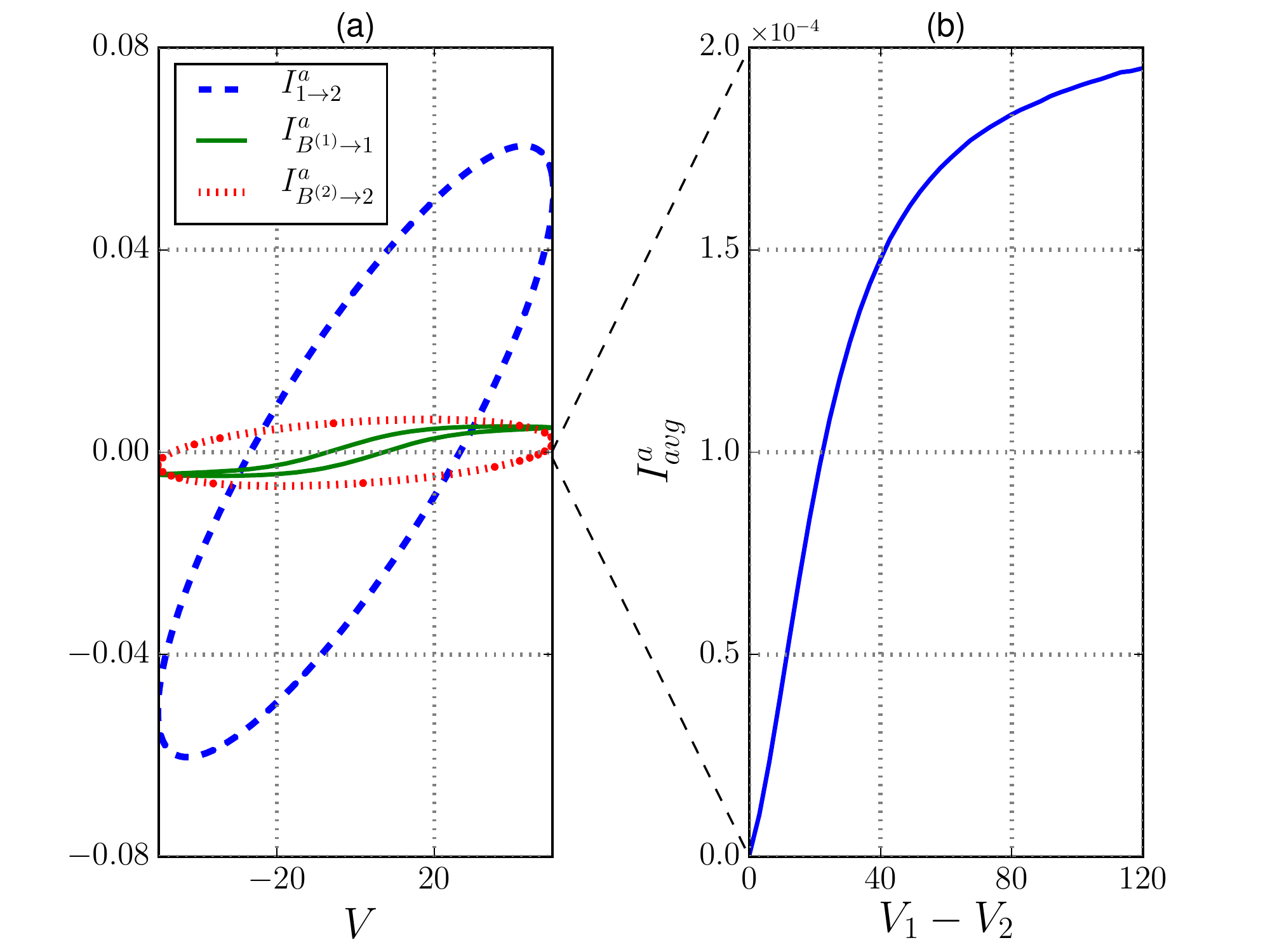} 
\caption{Panel (a) shows long time results of currents vs voltage $V=\mu_1-\mu_2$ over one time period of ac drive ($V_1=50,V_2=5$). The system is driven by $\mu_1~=V_1\cos(\Omega_0 t)$, $\mu_2~=~V_2\sin(\Omega_0 t)$, system-bath coupling is asymmetric ($\Gamma_1=0.01,\Gamma_2=0.09$) and system is at first resonance ($\Omega_0=\omega_2-\omega_1=2g$). The currents show hysteretic behavior, due to the dynamical nature of the charge transport in and out of the system. Because of the resonance, the current inside the system is much greater than currents from the baths. Panel (b) shows the time-period-averaged current as a function of difference between amplitudes of drives $V_1-V_2$. ($V_2=5$, $V_1$ is varied.) The current increases first and then reaches a plateau. The averaged current is much smaller than the maximum instantaneous current. The dotted lines between the two panels highlights the difference in scale of the plots. Other parameters $g=0.5,t_B=200,\omega_0=1,\beta_1=\beta_2=0.1$. All times are measured in units of $\omega_0^{-1}$, all energies are measured in units of $\hbar \omega_0$.} \label{fig:asym}
\end{figure}

\subsubsection{Asymmetric ac drive}

Having discussed the physics of long time dynamics of the symmetric ac drive, we move to the case of the asymmetric ac drive.
Consider the situation where the left chemical potential and right chemical potentials vary sinusoidally out of phase with same frequency but have different amplitude in general.

Unlike the symmetric ac drive case, here the voltage across the system $V=\mu_1-\mu_2$ is non-zero. For $V_1\neq V_2$. The difference between time-period averaged Fermi distributions of the two baths is not zero in this case. Hence, there is a net time-period averaged current through the system. Fig.~\ref{fig:asym} shows plots of currents from the baths and current in the system at first resonance. Panel (a) shows current vs voltage ($V=\mu_1-\mu_2$) curve over one time period. We observe an interesting hysteresis behavior. Also, because of resonance, the current inside the system is much greater than current from the baths. This effect would not have been seen in the adiabatic limit. Panel (b) shows the time-period averaged current as a function of $V_1-V_2$. The current increases, and finally saturates (similar to $I_{max}$ vs $V_0$ curve for the symmetric drive case).  Note that the time-averaged current is much smaller than the maximal instantaneous currents. This implies that perhaps time-dependent signals can be measured even when the average currents are small and below the noise level. 

The hysteresis in I-V curve for asymmetric system bath coupling can potentially have device applications. This behavior depends intricately on the phase difference between the two drives, the amplitude difference of the two drives, as well as on the asymmetry of system-bath coupling. These dependences are quite complicated and a detailed investigation of the hysteresis behavior will be taken up in a future work. 

The two site set-up described above is experimentally realizable using quantum dots. However, the bottleneck experimental parameter for observation of most of the above effects is frequency of the drive $\Omega_0$. Our most interesting results are close to resonance $\Omega_0=2g$. The maximum frequency of ac drive currently experimentally realizable is $\sim 10 GHz$ \cite{natcomm2016}. This means to observe the above effects $g\sim 0.01 meV$. The baths must have much wider bandwidths than system energy scales, which can be easily arranged experimentally. As evidenced by Eq.~\ref{Jform}, having wide bandwidths will automatically realize the weak system-bath coupling. Temperature $\sim 1mK$ (which is equivalent to $\beta\sim 0.1$) will be consistent with the Markov approximation.  These experimental parameters may be challenging but not impossible.  Our study thus points to new rich physics in such experimental domain.

\subsection{Conclusion}

We have formulated a theory of an out-of-equilibrium open quantum system set-up where the system is connected to multiple baths and is driven  by periodically varying thermodynamic parameters (temperature or chemical potential) of the baths. The key assumptions in our theory are weak system-bath coupling (Born approx) and that the relaxation time scale of the bath is much smaller than the time period of the drive (Markov approx). We do not make any assumption on the relaxation time scales of the system. This has allowed us to go beyond the adiabatic approximation where the system is always assumed to be in the steady state with the instantaneous bath. Our theory works for systems of non-interacting particles (i.e, system Hamiltonian is quadratic) bilinearly coupled to multiple baths (also quadratic Hamiltonian) in a lattice of any dimension or geometry (the particles may be all bosonic or all fermionic, boson-fermion couplings are not bilinear). Even though there exists formally exact methods of treating such periodically driven microscopic models, they are generally quite difficult to calculate beyond the adiabatic regime. On the other hand, phenomenological models based on Lindblad equations have been shown to be either quite restrictive or to have severe shortcomings. Unlike these, under Born-Markov approximation, our theory gives a much simpler, more transparent and much less restrictive results, providing a complete set of linear differential equations for equal time two-point correlation functions from which various physical observables (for example currents and populations) can be directly obtained. Also various physical effects can be directly read-off from such evolution equation for correlation functions. The long time dynamics of the system are readily shown to be periodic with the same period as the drive. The occurrence of resonance when the frequency of the periodic drive equals the difference of two eigen-energies of the system is also easily seen from the equations. 

We have tested our theory by applying it to the case of two fermionic sites with hopping between them, weakly coupled to two different baths at same constant temperature but with sinusoidally driven chemical potentials. In this we have considered two cases. We have validated our theory by comparing with exact numerics done with finite but large baths for both the cases. The first case is when the sinusoidal drive is symmetric, i.e, both chemical potentials vary in an exactly same manner so that there is no instantaneous voltage difference. In this case, even though there is no net time-period averaged current, there is an instantaneous current, which can become quite large at resonance. Furthermore, inside the system, the instantaneous current depends on asymmetry of the system-bath coupling and becomes zero when system-bath coupling is symmetric. This gives an experimental way of detecting the asymmetry of system-bath coupling. The maximum instantaneous current increases with amplitude of drive and shows resonance peaks. Increase in temperature washes out the higher resonance peaks.

In the other case, the chemical potentials of the baths have an amplitude difference, as well as a phase difference. Due to amplitude difference, the net time-period averaged current is non-zero and increases with increase in amplitude difference. We find interesting hysteresis in the I-V curves, a detailed investigation of which will be taken up in a future work.

Two most important experimentally relevant insights from our results, especially for studying electronic transport in molecular junctions or through quantum dots, are : (i) time-dependent measurements of current for symmetric  oscillating voltages (with zero instantaneous voltage bias) can point to the degree of asymmetry in the system, and (ii) under certain conditions, time-dependent currents can exceed time-averaged currents by several orders of magnitude, and can therefore be detected even when the average current is below the measurement threshold.

Since the formalism is for non-interacting fermions (bosons) on a lattice, bilinearly coupled to fermionic (bosonic) baths, interactions can be built in at a mean field level. Thus our theory can be used to describe a wide range of experimental set-ups (such as molecular junctions, quantum dots, cavity-QED expts, cold atoms, cavity optomechanics etc.). However, the theory cannot be applied to extremely low temperatures, because of the Markov approximation. The relaxation time scale of a bath can be shown to be inversely proportional to the temperature of the bath. Hence, at extremely low temperatures, the Markov approximation would be violated. Also, our formalism cannot treat cases where system-bath coupling is not bilinear, such as fermions connected to bosonic baths and vice-versa. Further work is required for go beyond these limitations.  Other future work includes treating interacting systems beyond mean-field, as well as incorporating the effects of dephasing or local vibrations.    
\vspace{100pt}

\paragraph*{Acknowledgements: } Y.D. acknowledges support from the Israel Science Fund (Grant No. 1256/14). A.P would like to thank Prof Abhishek Dhar for very insightful discussions, and also for funding support from his grant under the Indo-Israel joint research project No.6-8/2014(IC).

\appendix

\section{Derivation of RQME for time independent case}\label{Appendix_A}
The Redfield equation for our set-up in the dc bias case has been rigorously deriven in Ref \cite{Purkayastha2016}. We include the same derivation here for completion.

The standard Redfield equation is derived assuming weak-system bath coupling (Born approximation) \cite{CarmichaelBook,breuer2007theory}. 
For bilinear system-bath coupling $\hat{\mathcal{H}}_{SB} = \sum_{\ell} \hat{S}_\ell \hat{B}_\ell$
where $\hat{S}_\ell$ operates on system and $\hat{B}_\ell$ operates on bath, system-bath coupling being turned on at time $t=0$, we have the master equation  \cite{CarmichaelBook}
\begin{align}
\label{gen_master}
\frac{\partial\rho^I}{\partial t} &= -\sum_{\ell,m} \int_0^t dt^{\prime} \Big\{ [\hat{S}^I_\ell(t),\hat{S}^I_m(t^{\prime})\rho^I(t)]\langle \hat{B}^I_\ell(t)\hat{B}^I_m(t^\prime)\rangle_B  \nonumber \\ & +[\rho^I(t)\hat{S}^I_m(t^{\prime}),\hat{S}^I_\ell(t)]\langle \hat{B}^I_m(t^\prime)\hat{B}^I_\ell(t)\rangle_B \Big\} 
\end{align}
where we use the interaction representation $\hat{O}^I(t)=e^{i(\hat{\mathcal{H}}_S+\hat{\mathcal{H}}_B)t} \hat{O} e^{-i(\hat{\mathcal{H}}_S+\hat{\mathcal{H}}_B)t}$ and  $\langle...\rangle_B$ refers to the average taken only with respect to bath.  For our set-up in Eq.~\ref{model_H}, we have the corresponding system and bath operators in interaction picture $\hat{a}_{\ell}^{I}(t) = \sum_{\alpha=1}^N c_{\ell \alpha} \hat{A}_\alpha e^{-\omega_\alpha t}$, and $\hat{B}_{\ell r}^{I}(t) =  \hat{B}_{\ell r} e^{-\Omega_{\ell r}t}$, where we have used definitions in Eq.~\ref{Aop}. Putting these in Eq.~\ref{gen_master} and going back to Schroedinger picture, one obtains
\begin{equation}
\label{qme_deriv}
\begin{split}
&\frac{\partial\rho}{\partial t} = i[\rho, \hat{\mathcal{H}}_S] - \varepsilon^2\sum_{\alpha,\nu,\ell=1}^N c_{\ell \alpha}^* c_{\ell \nu}\int_{\Omega_{min}^{\ell}}^{\Omega_{max}^{\ell}} \frac{d\omega}{2\pi} \Big[ \Big\{[\rho(t) \hat{A}_\nu, \hat{A}_\alpha^{\dagger}]  \\
&+[\hat{A}_\alpha^{\dagger}, \hat{A}_\nu \rho(t)] e^{\beta_\ell(\omega-\mu_\ell)} \Big\} J_\ell(\omega) n_\ell(\omega)  \int_0^t d\tau e^{i(\omega - \omega_\nu)\tau}  \\&+\hspace{2pt}  h.c. \Big]
\end{split}
\end{equation}
where $\Omega_{min}^{\ell},\Omega_{max}^{\ell}$ are the minimum and the maximum energy levels of the bath coupled to the $\ell$th site. This equation still has an integration over time from $0$ to observation time $t$ and hence has time-dependent rates. To obtain a QME with time independent rates, one assumes $t \gg \tau_B$ (Markov approximation). This  allows one to take the $t \rightarrow \infty$ in the integration limit. Upon doing this, we obtain the microscopically derived Born-Markov RQME with time independent rates. Multiplying this equation by $\hat{A}_{\alpha}^{\dagger}\hat{A}_{\nu}$ and taking trace, one can obtain the closed set of linear differential equations for the equal time two point correlation functions. This equation will be exactly same as Eq.~\ref{C_differential} except that $Q$  will be time independent.

\section{Finding $\tau_B$}\label{Appendix_B}

\begin{figure}
\includegraphics[width=\columnwidth]{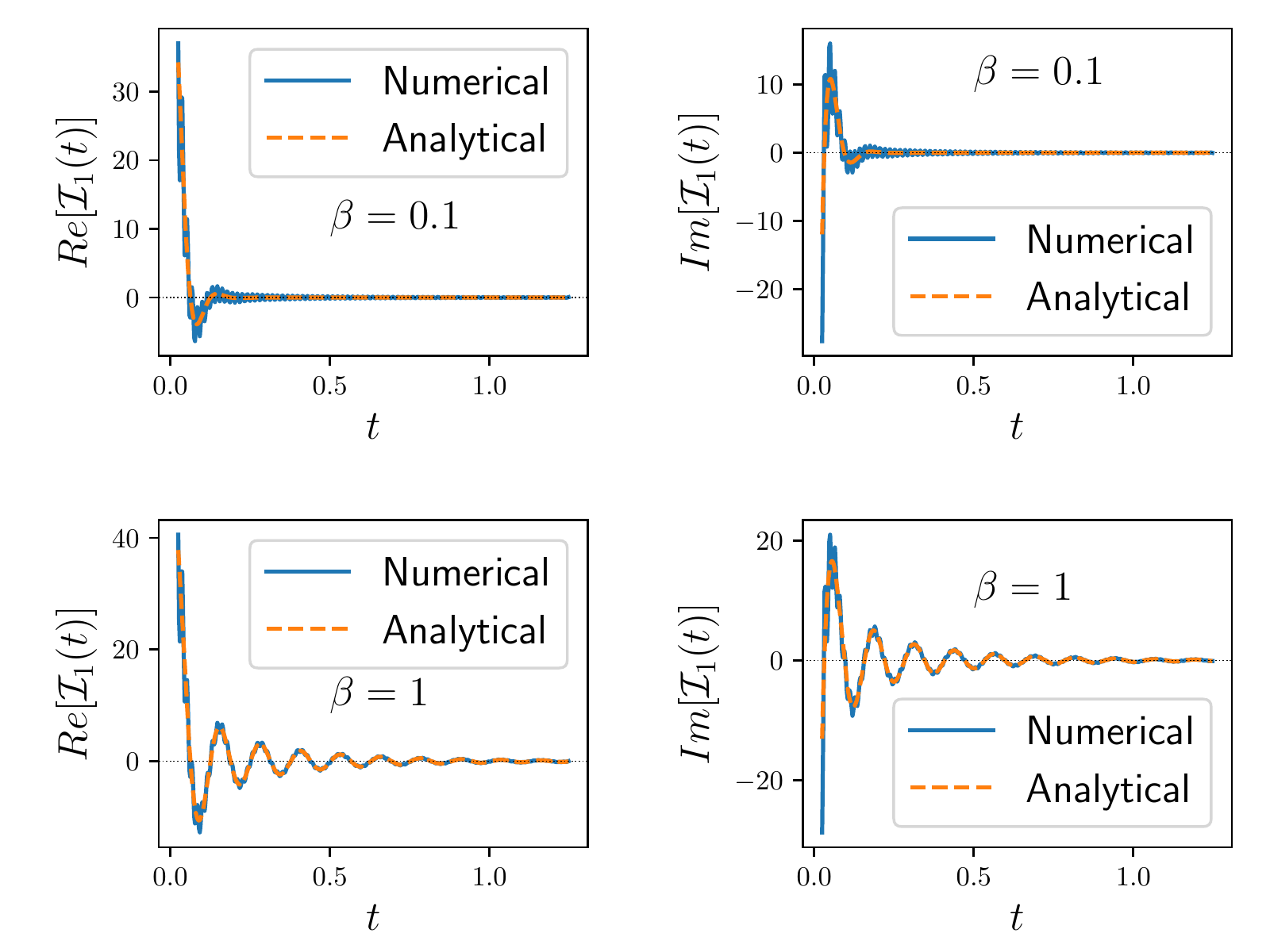} 
\caption{\textbf{Temeprature dependence of $\tau_B$:} The figure shows plots of real and imaginary $\mathcal{I}_1(t)$ (Eq.~\ref{I1}), calculated via exact numerical integration and via the analytical result Eq.~\ref{I1_analytical} for inverse temperature $\beta=0.1$ (top panels), and $\beta=1$ (bottom panels). The numerical and analytical results agree perfectly. The decay time for $\mathcal{I}_1(t)$ gives the measure of bath relaxation time $\tau_B$. The temperature dependence of $\tau_B$ is clear from the plots. From Eq.~\ref{I1_analytical}, it is clear that $\tau_B\sim \beta$. Other parameters: $t_B=200,\mu=50$. All times are measured in units of $\omega_0^{-1}$, all energies are measured in units of $\hbar \omega_0$.} \label{fig:tau_B}
\end{figure}

The condition on time scales Eq.~\ref{t_cond1} is one of the major requirements for validity of our theory. This entails that we must find a good estimate of bath relaxation time $\tau_B$. We have seen above that to derive the  RQME with time independent rates, we need observation time $t \gg \tau_B$, so that we can take upper limits of certain integrals to infinity. We can thus look more carefully at the integrals to check under what conditions of bath parameters such an assumption is valid. Looking at Eq.~\ref{qme_deriv}, and using the spectral function in Eq.~\ref{Jform} and fermi distribution, we find that the relevant integrals are $\int_0^t \mathcal{I}_1(t^\prime) dt^\prime$ and $\int_0^t \mathcal{I}_2(t^\prime) dt^\prime$, where
\begin{align}
&\mathcal{I}_1(t) = \frac{2\gamma^2}{t_B}\int_{-2t_B}^{2t_B} \frac{d\omega}{2\pi}
\sqrt{1-(\frac{\omega}{2t_B})^2}\frac{1}{e^{\beta(\omega-\mu)}+1}e^{i\omega t} \label{I1}\\
&\mathcal{I}_2(t) = \frac{2\gamma^2}{t_B}\int_{-2t_B}^{2t_B} \frac{d\omega}{2\pi}
\sqrt{1-(\frac{\omega}{2t_B})^2}\frac{e^{\beta(\omega-\mu)}}{e^{\beta(\omega-\mu)}+1}e^{i\omega t} \label{I2}
\end{align}
The bath relaxation time scale $\tau_B$ is essentially the time in which $\mathcal{I}_1(t)$ and $\mathcal{I}_2(t)$ decay. If the observation time is much larger than $\tau_B$, while doing the time integrals, the upper limit can be extended to infinity, because, the contribution to the time integral from time greater than $\tau_B$ will be negligible. Using contour integrals, $\mathcal{I}_1(t)$ and $\mathcal{I}_2(t)$ can be analytically evaluated. If $t \gg (2t_B)^{-1},(2t_B-\mu)^{-1}$, the leading order term for $\mathcal{I}_1(t)$ is given by
\begin{align}
\label{I1_analytical}
\mathcal{I}_1(t)\simeq & -\frac{2\gamma^2 i}{t_B\beta} e^{i\mu t}\sum_{0\leq r \leq r^*} \Big[
e^{-(2r+1)\pi t/\beta} \nonumber\\
& \sqrt{1-\frac{(2r+1)\pi i + \beta \mu}{2t_B \beta}} \Big]
\end{align} 
where $r$ are integers and $r^* = \frac{\beta}{2\pi}(2t_B - \mu) - \frac{1}{2}$. The validity of this result is shown in Fig.~\ref{fig:tau_B}, where $\mathcal{I}_1(t)$ calculated via exact numerical integration and via above equation are plotted for two different temperatures. Similar result with opposite sign holds for $\mathcal{I}_2(t)$. From this result, it is clear that, if $2t_B$ is quite large, then $\tau_B \sim \beta$, and is independent of $\mu$. However, in our AC voltage drive case, we want to vary $\mu$ as a function of time, and make the Markov approximation always. So, in our case, $\tau_B \sim \beta$. This means, as mentioned in the main text, the condition on time scales (Eq.~\ref{t_cond1}) render our theory inapplicable at very low temperatures.

We would like to mention that while this result was calculated explicitly for our chosen spectral function and fermi distribution, similar results yielding $\tau_B \sim \beta$  can also be obtained for other popular choices of spectral functions (for example, $J(\omega)~=~const$, $J(\omega)~\propto~\frac{\omega}{\omega^2 + \Lambda^2}$), as well as for bose distribution.

\section{Deriving Eq.~\ref{C_differential} rigorously from the protocol}\label{Appendix_C}
Now, we would like to demonstrate that, following through our protocol in Section~\ref{B} rigorously using the time-independent RQME leads to Eq.~\ref{C_differential}. Between the $(r-1)$th and the $r$th steps of the protocol, the evolution equation for correlation matrix is as given by time-independent RQME
\begin{align}
\frac{dC((r-1)\tau_D}{dt} = -MC((r-1)\tau_D) + \varepsilon^2 Q ((r-1)\tau_D)
\end{align}
The correlation matrix at time $r\tau_D$ is then given by the formal solution of this equation which can be written as
\begin{align}
e^{M\tau_D}C(r\tau_D) =& C((r-1)\tau_D) \nonumber\\
&+ \varepsilon^2\int_0^\tau dt^\prime e^{Mt^\prime} Q ((r-1)\tau_D+t^\prime)\label{formal}
\end{align}
If $\tau_D$ is small, $t^\prime$ is also small. But $r\tau_D$ can still be large for $r\gg 1$. In such case, we can expand each term to the linear order in $\tau_D$ to obtain
\begin{align}
&(1+M\tau_D)C(r\tau_D) = C((r-1)\tau_D)+ \varepsilon^2  Q ((r-1)\tau_D) \tau_D \nonumber \\
&\Rightarrow\frac{C(r\tau_D)-C((r-1)\tau_D)}{\tau_D} = -MC(r\tau_D) \nonumber\\
&\hspace{125pt}+ \varepsilon^2 Q ((r-1)\tau_D)
\end{align}
This equation is the discrete time version of Eq.~\ref{C_differential} and for $\tau_D \rightarrow 0$ and $r\tau_D \rightarrow t$, exactly gives Eq.~\ref{C_differential}.

Eq.~\ref{C_differential} may also be obtained by `putting by hand' time dependence in the fermi or bose distributions that appear in the dc bias RQME. However, note that this form would not have been possible without the Markov approx $\tau_{D} \gg \tau_B$. Without Markov assumption, the matrix $M$ would be time dependent which will not lead to a formal solution of the form Eq.~\ref{formal}, and as a consequence would not lead to Eq.~\ref{C_differential}.

\section{Derivation of spectral function of bath Eq.~\ref{Jform}}\label{Appendix_D}

We now derive the bath spectral function Eq.~\ref{Jform} for the explicit bath Hamiltonian in Eq.~\ref{ham2S}. In Eq.~\ref{model_H}, the system couples to eigenmodes of the bath. Hence we can obtain the spectral function by using the eigen-energies and the eigenvectors of the bath. Since the bath has infinite number of modes, the spectrum can be assumed continuous. With this, the bath eigen-energies  are $\Omega(q_\ell) = -2t_B \cos q_\ell$ and system-bath coupling to bath modes are given by $\kappa(q_\ell)=\gamma_\ell\sqrt{\frac{2}{\pi}}\sin q_\ell$, with $0 \leq q_\ell \leq \pi$. Thus, 
\begin{equation}
\begin{split}
J_{\ell}(\omega) =& 4\gamma_\ell\int_0^\pi dq_\ell \sin^2 q_\ell \hspace{3pt}\delta(\omega+2t_B \cos q_\ell) \\
& = \frac{2\gamma_\ell^2}{t_B} \sqrt{1-\frac{\omega^2}{4t_B^2}}
\end{split}
\end{equation}
We also need to the following result $ \mathcal{P} \int \frac{}{}\frac{d\omega^{\prime} J_{\ell}(\omega^{\prime})}{2\pi(\omega^{\prime}-\omega)}=-\frac{\gamma_\ell^2\omega}{2t_B^2}$ to calculate $f_{\alpha\nu}^{\Delta}(\omega)$ [in Eq.~ \ref{C_differential}]. $F_{\alpha\nu}^{\Delta}(\omega)$ cannot be written in a simple closed form and is calculated numerically.

\bibliography{ac_drive_citations_prb}

\end{document}